\documentclass[twocolumn,prl]{revtex4-1}
\pdfoutput=1

\usepackage{amsmath}
\usepackage{graphicx}
\usepackage{xcolor}
\usepackage{bm}

\newcommand{\eq}{\begin{equation}}
\newcommand{\eeq}{\end{equation}}
\newcommand{\aeq}{\begin{equation}\begin{aligned}}
\newcommand{\eaeq}{\end{aligned}\end{equation}}

\newcommand{\Qij}{Q_{ij}}
\newcommand{\vi}{v_i}

\newcommand{\blue}[1]{{\color{black} #1}}

\newcommand{\red}[1]{{\color{black} #1}}

\newcommand{\rev}[1]{{\color{black} #1}}

\begin{document}

\title{Activity suppressed phase separation}

\author{Fernando Caballero}
\email{fmc36@ucsb.edu}
\affiliation{Department of Physics, University of California Santa Barbara, Santa Barbara, CA 93106}
\author{M. Cristina Marchetti}
\affiliation{Department of Physics, University of California Santa Barbara, Santa Barbara, CA 93106}
\date{\today}

\begin{abstract}
    We use a continuum model to examine the effect of activity on a phase separating mixture of  an extensile active nematic and a passive fluid. We highlight the distinct role of \red{(i)} previously considered interfacial active stresses and \red{(ii)} bulk active stresses that couple to liquid crystalline degrees of freedom. Interfacial active stresses can arrest phase separation, as previously demonstrated. Bulk extensile active stresses can additionally strongly   suppress phase separation by sustained self-stirring of the fluid, substantially reducing the size of the coexistence region in the temperature/concentration plane relative to that of the passive system.  The phase separated state is a dynamical emulsion-like steady state of continuously splitting and merging droplets, as suggested by recent experiments. Using scaling analysis and simulations, we identify various regimes for the dependence of droplet size on activity. These results can provide a criterion for identifying the mechanisms responsible for arresting phase separation in experiments.
\end{abstract}
\maketitle

Liquid-liquid phase separation (LLPS) occurs ubiquitously in biology and materials science. In this process two immiscible fluids demix from a homogeneous state into two distinct liquids, separated by soft interfaces.   Recent attention has focused on  LLPS  in active systems~\cite{ray_unpublished,Doost2018}. Persistent motility is known to drive phase separation in systems in which the constituent particles have purely repulsive interactions through a process known as motility induced phase separation (MIPS)~\cite{tailleur2008statistical,Fily2012, redner2013}. \blue{This is in contrast to equilibrium phase separation, which requires attractive interactions and occurs when attraction overcomes thermally-driven entropic mixing.} MIPS has been characterized extensively, experimentally and theoretically, resulting in a detailed understanding of its underlying mechanisms~\cite{MIPS}, phases~\cite{C3SM52813H,C3SM52469H}, critical properties~\cite{Caballero_2018}, and laboratory realizations~\cite{doi:10.1126/science.1230020, doi:10.1126/science.1140414}.

Less \red{studied} is the effect of activity on the phase diagram of fluid mixtures that do phase separate in equilibrium. 
Very recent experiments on immiscible phase separating mixtures of active and passive fluids~\cite{Tayar2021} have suggested \red{that activity may both \emph{arrest} phase separation, i.e., stabilize finite-size structures, preventing the system from reaching bulk phase separation, and also \blue{\emph{suppress} it} by shifting the 
critical temperature to lower values and reducing the region of the  phase diagram where the two fluids are demixed.} 
While binary mixtures of active and passive species have been studied and modeled before in different contexts~\cite{Blow2014,Hughes2020,Giomi2014,williams2022confinement,xu2021morph}, an understanding of this suppression is still missing.

Equilibrium \rev{liquid-liquid} phase separation is well described by a continuum theory in terms of a conserved \red{concentration} field $\phi(\mathbf{r},t)$ with an underlying free energy quartic in $\phi$ \rev{coupled to a flow with velocity field $\mathbf{v}$}~\cite{chaikin1995principles}. 
\red{Activity has been introduced in  equilibrium models through the addition of \rev{time reversal symmetry breaking} terms to the currents that drive the dynamics of the concentration field~\cite{Wittkowski2014} and through interfacial stresses that describe activity-induced self-shearing of the interface~\cite{Tiribocchi2015,Singh2019}. It has been shown that these nonequilibrium mechanisms can arrest phase separation and stabilize finite-size droplets~\cite{Singh2019,Tjhung2018}, but, as we see below, do not change the critical temperature of the passive phase separating mixture.}

In this Letter we examine the effect of activity on a phase separating mixture of a passive fluid and an active nematic and show that in this system activity not only arrests phase separation, but can completely suppress it.
The suppression originates from the presence of liquid crystalline degrees of freedom that generate \emph{bulk} active stresses not present in scalar models.
It is known that such active stresses destabilize the homogeneous nematic state on all scales~\cite{PhysRevLett.89.058101}, resulting in turbulent-like dynamics. \blue{This yields a  continuous local stirring that mixes} the fluid into a uniform state, shifting the critical point for phase separation to lower temperatures and reducing the coexistence region in the temperature/concentration plane.
A similar suppression is known to occur in equilibrium LLPS upon imposing a uniform external shear~\cite{Han2006,Onuki1997,Onuki1995,Silberberg1952}. The same effect is achieved here via local \emph{self-shearing}, demonstrating that activity provides a new handle for controlling the LLPS phase diagram.

\emph{Model.} 
We consider a phase separating mixture of a passive isotropic fluid and an active nematic~\footnote{\blue{The model can be adapted to describe a mixture of active and passive nematic by decoupling the $Tr[\mathbf{Q}^2]$ term in the free energy from $\phi$, while maintaining the coupling to $\phi$ in the active stress.}}. Such a system has recently been engineered by combining active microtubule nematics with DNA-based condensates~\cite{Tayar2021}. We work in two dimensions ($2D$), although the analytical work is easily generalized to $3D$. The system is described by three continuum fields: a phase field $\phi=\langle n_A-n_P\rangle/\langle n_A+n_P\rangle$ that represents the local composition of the fluid mixture, with $n_{A,P}$ the number density of active and passive particles, a flow velocity $\mathbf{v}$, and the nematic order parameter tensor $\Qij=S/2(n_in_j-\delta_{ij}/2)$, where $S$ is the amplitude and $\mathbf{n}$  the director, with $|\mathbf{n}|=1$. The dynamics is governed by 
\aeq\label{eq:motion}
D_t\phi &= M\nabla^2\mu\;,\\
D_t Q_{ij} &= \lambda D_{ij} + Q_{ik}\omega_{kj}-\omega_{ik}Q_{kj} - \frac{1}{\gamma}\frac{\delta\mathcal F_Q}{\delta Q_{ij}}\;,\\
\rho D_t v_i &= \eta\nabla^2 v_i -\partial_i P +f_i\;,
\eaeq
with $D_t=\partial_t+\vi\partial_i$. The field $\phi$ obeys Cahn-Hilliard dynamics with a mobility $M$ and a chemical potential $\mu=\frac{\delta\mathcal F_{\phi}}{\delta\phi}$, obtained from a Landau-Ginzburg \red{(LG)} free energy 
$\mathcal F_\phi = \frac{1}{2}\int_\mathbf{r}\left[ a\phi^2+\frac{b}{2}\phi^4 + \kappa(\nabla\phi)^2\right]$.
The parameter $a$ represents the temperature of the passive system and controls the equilibrium critical point $a_c$ located at $a_c=0$ for $\phi_0=\int_{\mathbf{r}}\phi(\mathbf{r})/A=0$, with $A$  the system area. Considering states of uniform $\phi(\mathbf{r})=\overline{\phi}$, for $a>0$ the free energy $\mathcal F_\phi$ has a single minimum at $\overline{\phi}=0$, corresponding to a uniform (mixed) state. For $a<0$ the system minimizes the free energy by demixing into two bulk coexisting states with \red{$\overline{\phi}=\pm\phi_+=\pm\sqrt{-a/b}$}. For convenience we also define  $\tilde\phi=(1+\phi/\phi_+)/2$, so that $\tilde\phi=1$ corresponds to the active nematic phase and $\tilde\phi=0$ to the isotropic passive fluid. 

The dynamics of the nematic order parameter is controlled by coupling to flow through the symmetrized rate of strain tensor $D_{ij}=(\partial_iv_j+\partial_jv_i)/2$, vorticity $\omega\red{_{ij}}=(\partial_iv_j-\partial_jv_i)/2$, and relaxation controlled by the Landau-deGennes free energy $\mathcal F_Q$, given by
\eq\label{eq:FQ}
\mathcal F_Q = \frac{1}{2}\int_\mathbf{r}\left[ Tr[\mathbf{Q}^2]\left( Tr[\mathbf{Q}^2]- \tilde\phi\right) + K(\partial_jQ_{ik})^2\right]\;.
\eeq
The coupling to $\tilde\phi$ ensures that nematic order does not occur in the passive region, where $\tilde\phi=0$.

The flow is governed by a Navier-Stokes equation, with $\rho$  the constant total density, $\eta$ the shear viscosity and $P$ the pressure, fixed by incompressibility, $\bm\nabla\cdot\mathbf{v}=0$. The force density $\mathbf{f}$ has two contributions, $\mathbf{f}=\mathbf{f}^\phi+\mathbf{f}^Q$. The  capillary force,  $\mathbf{f}^\phi = -\phi\bm\nabla \mu$~\cite{Bray2002}, can be written as the divergence of a deviatoric stress, $\mathbf{f}_i^\phi=\partial_j\sigma_{ij}^\red{\phi}=-\kappa_a\partial_j[(\partial_i\phi)(\partial_j\phi)-\delta_{ij}(\nabla\phi)^2/2]$~\cite{Cates2018}. We allow for $\kappa_a$ to be different from the stiffness $\kappa$ in the \red{LG} free energy  to describe non equilibrium interfacial stresses, akin to active model H~\cite{Tiribocchi2015}. The sign of $\kappa_a$ controls the nature of interfacial active stresses, with $\kappa_a>0$ and $\kappa_a<0$ corresponding to stresses that tend to stretch and contract the interface along its length, respectively (see Fig. SM1). As shown previously~\cite{Tiribocchi2015,Singh2019}, $\kappa_a>0$ stabilizes the interface, while $\kappa_a<0$ results in an effective negative interfacial tension and destabilizes it.
The bulk active force $\mathbf{f}^{Q}$ is controlled by the liquid crystalline degrees of freedom and is the gradient of the familiar active stress $f_i^{Q}=\partial_j\sigma_{ij}^a=\partial_j(\alpha\tilde\phi\Qij)$~\cite{Marchetti2013}.
The coupling to $\tilde\phi$ ensures that the active stress is nonzero only in the nematic phase.
The sign of $\alpha$ controls whether bulk active stresses are contractile ($\alpha>0$) or extensile ($\alpha < 0$). 
Related models have been used before to describe coexisting nematic and isotropic phases of active liquid crystals~\cite{Blow2014,Hughes2020,Giomi2014,coelho2022dispersion}, but the  boundary of the coexistence region has not been previously quantified.

\begin{figure}
    \includegraphics[scale=0.18]{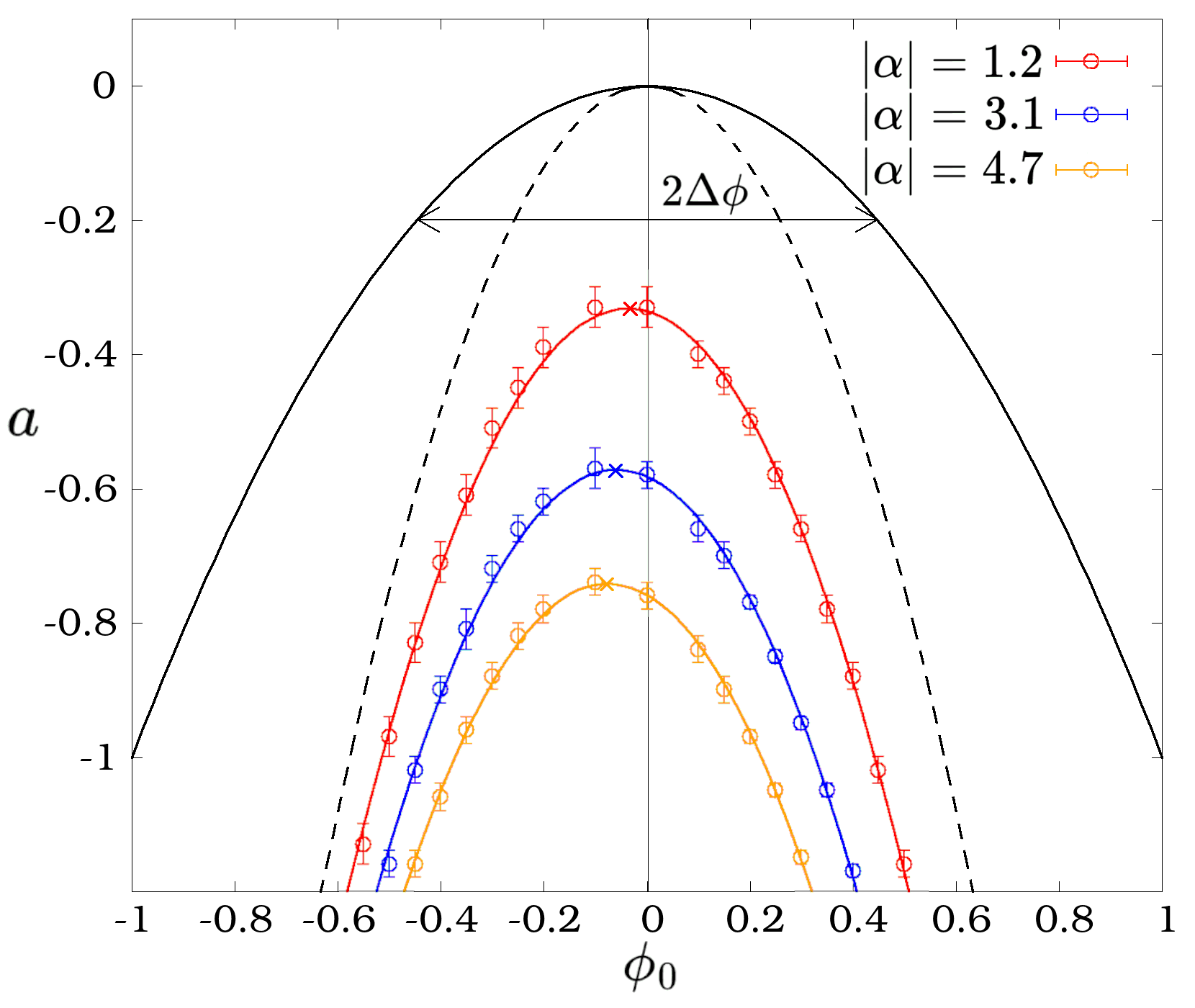}
    \caption{\label{fig:pl2d}
    Phase diagram in the $(\phi_0,\alpha)$ plane.
   The black lines are the binodal (solid) and spinodal (dashed) curves for the mean field theory in equilibrium. 
   The numerical data for three values of activity $\alpha$ clearly show the shift of the critical point and its increase with activity. The lines through the data are parabolic fits and the thick dots mark the critical point. The vertical line at $\phi_0=0$ highlights the shift \blue{of the critical point to values of $\phi_0$ that correspond to more than $50\%$ of passive fluid.} The error bars are the precision with which we can visually differentiate uniform and phase separated states.}
\end{figure}

\emph{Numerical Results.} 
In equilibrium ($\kappa_a=\kappa$ and $\alpha=0$), our fluid mixture undergoes bulk phase separation for $a<0$. The binodal (coexistence)  curve  $a=-\phi_0^2$ and the spinodal curve $a=-3\phi_0^2$ (we set $b=1$) are symmetric inverted parabolas shown in Fig.~\ref{fig:pl2d} as solid and dashed black lines, respectively. Below the spinodal parabola, an initially uniform state is unstable to small density fluctuations and phase separates through spinodal decomposition. Between the binodal and spinodal lines, uniform states are metastable, and phase separate through nucleation. Above the binodal line, the uniform state is stable. 

\emph{Interfacial} active stresses alone ($\kappa_a<0$ and $\alpha=0$) narrow the  coexistence region, changing the coexistence densities and arresting phase separation, but do not affect the location of the critical point~\cite{Tiribocchi2015,SM}. In the following we focus on the effect of bulk activity and assume $\kappa_a=\kappa$ since the effect of $\kappa_a<0$ has been discussed elsewhere~\cite{Tiribocchi2015,Singh2019}

The situation  is completely different in the presence of extensile \emph{bulk} active stresses ($\alpha<0$ and $\kappa_a=\kappa$) \blue{that are known to destabilize the uniform ordered state of extensile active nematics for all $\alpha$~\cite{PhysRevLett.89.058101}.} \red{This} instability is driven by bend deformations that grow in time, leading to a dynamical steady state of spatio-temporal chaotic flows known as active turbulence~\cite{doi:10.1146/annurev-conmatphys-082321-035957}. To quantify the effect of the bulk instability on the phase diagram \red{of the mixture}, we integrate numerically Eqs.~\ref{eq:motion}. (See~\cite{SM} for details on the numerics).
We vary volume fraction $\phi_0$, reduced temperature $a$ and activity $\alpha$. The rest of the parameters are as follows, unless specified otherwise: $\eta=0.1,\gamma=1,K=0.08,\lambda=1,M=0.1,b=1,\kappa=\kappa_a=0.7$.

\begin{figure}
    \includegraphics[scale=0.18]{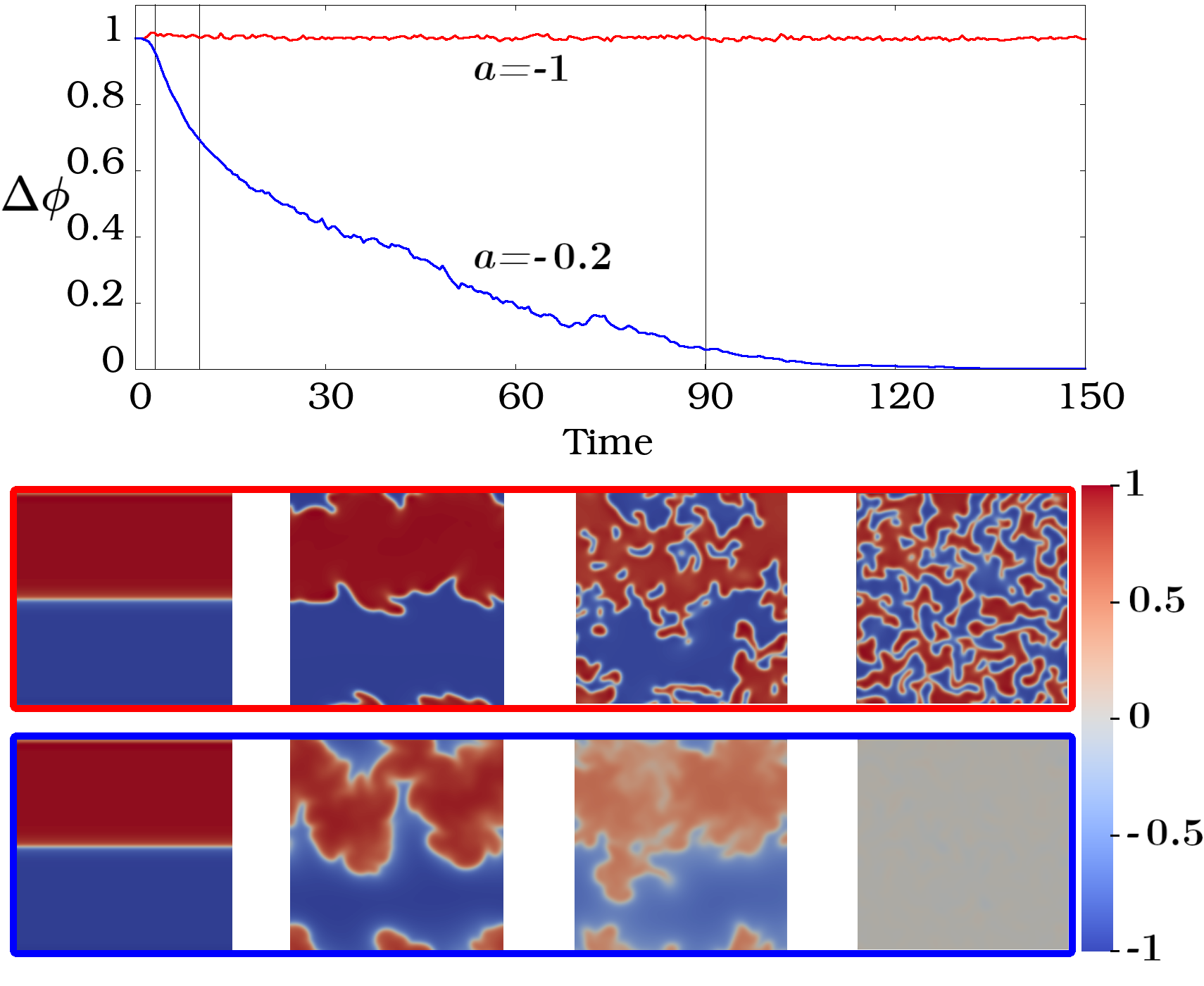}
    \caption{\label{fig:time_ev} Top: Time evolution of the width $\Delta\phi$ of the coexistence region for $a=-0.2$ (blue curve) and $a=-1.0$ (red curve), corresponding to states above and below the critical point $a_c^*\approx-0.3$ of the active mixture, respectively. Here $|\alpha|=0.8$. Bottom:  snapshots of the simulations at times $t=0,3,10$ and $90$ marked by vertical lines in the plot. The color of the frame of the snapshots matches the color of the corresponding $\Delta\phi$ curves.}
\end{figure}

We find that bulk extensile stresses not only arrest phase separation, but considerably suppress it, by shifting the critical point $a_c$ to a lower value $a_c^*(\alpha,\phi_0)<0$ and to $\phi_0<0$.   The area of the coexistence region in the $(\phi_0,a)$ plane is substantially reduced, as shown in Fig.~\ref{fig:pl2d}. To quantify this suppression, we examine the time evolution of the width of the coexistence region $\Delta\phi(a)=[\text{max}(\phi)-\text{min}(\phi)]/2$ (Fig.~\ref{fig:pl2d}).
In the passive system  $\Delta\phi(a)=\phi_+=\sqrt{-a}$, with $\Delta\phi=0$ in the uniform state.
The time evolution of $\Delta\phi$ is shown in the top frame of Fig.~\ref{fig:time_ev}  for two values of activity, both corresponding to $a<0$. For $a_c^*<a<0$, i.e., above the active critical temperature  ($a=-0.2$, blue curve),  $\Delta\phi$ quickly evolves to zero as the initially phase separated state is mixed by active flows. For $a<a_c^*$ ($a=-1$, red curve), the active fluid evolves towards arrested phase separation.
The boundaries of the coexistence region shown in Fig.~\ref{fig:pl2d} are identified as the points where $\Delta\phi$ approaches a finite value at late times. We note that this criterion does not distinguish between binodal and spinodal lines.

The active critical point $a_c^*(\alpha,\phi_0)$ depends on activity and volume fraction. It is shifted downward with increasing activity, which increases the rate at which the mixture is stirred by active energy injection.
The parabolic fit to the data in Fig.~\ref{fig:pl2d} shows that activity additionally shifts the critical point to lower (negative) values of $\phi_0$, which corresponds to mixtures with an excess fraction of active versus passive component.

\begin{figure}
    \includegraphics[scale=0.18]{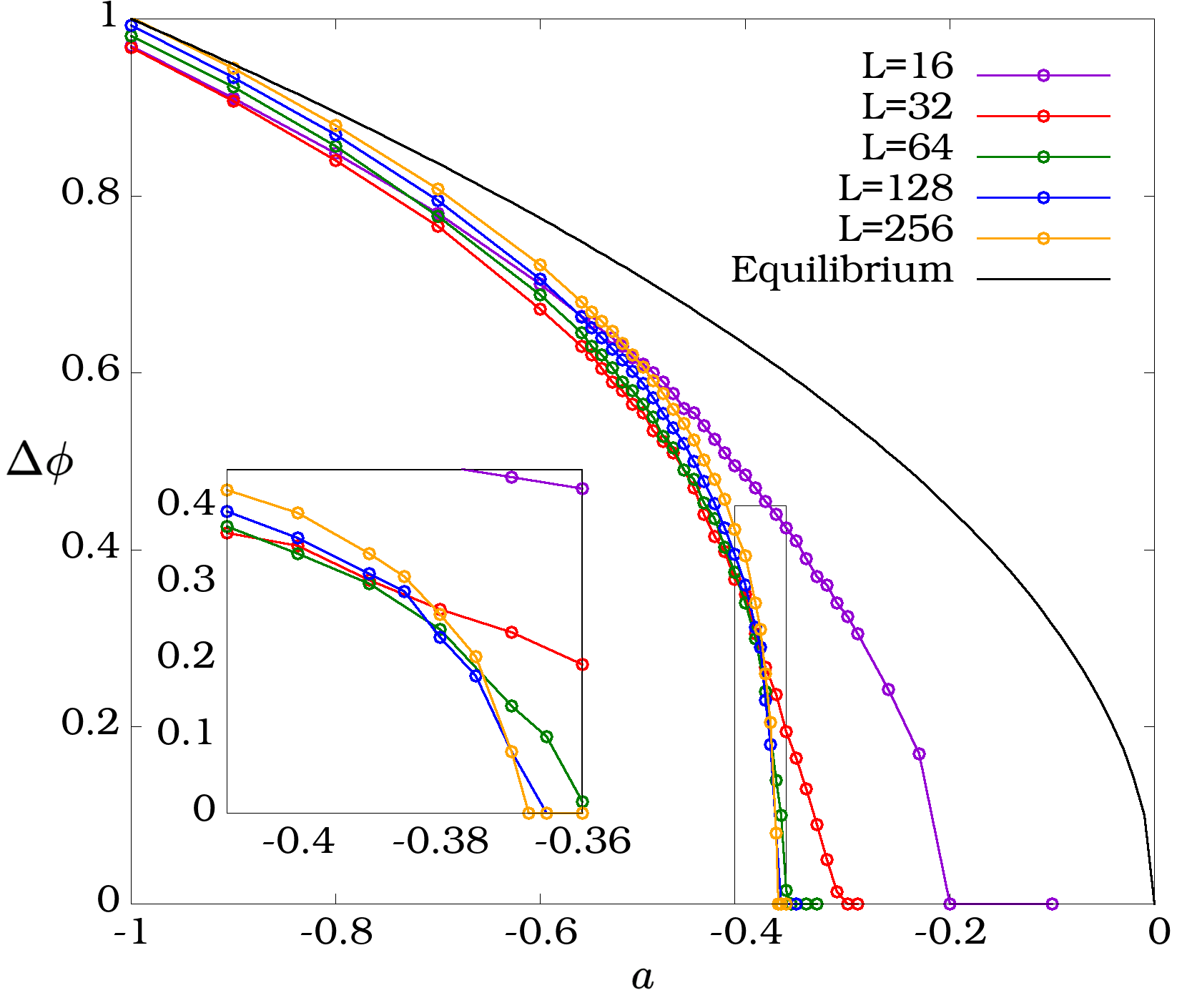}            
    \caption{\label{fig:fss} Finite size scaling  of $\Delta\phi(a,\phi_0^c)$ for $|\alpha|=1.2$. \blue{The critical point $a_c$ is defined as the value of $a$ where $\Delta\phi=0$.} The black line is the equilibrium mean field line of the passive system, \blue{with $a_c=0$.} \blue{For active mixtures $a_c$ is shifted to lower (negative) values and approaches 
    $a_c\approx-0.4$ for the largest system sizes considered. $d\Delta\phi/da$ diverges at the active critical point.} The inset shows a zoomed in portion of the data in the black rectangle.
    }
\end{figure}

To determine $a_c^*(\alpha,\phi_0)$, we first locate $\phi_0^c$ through a parabolic fit of the coexistence line, and then evaluate $\Delta\phi(a,\phi_0^c)$ as a function of $a$.
These curves are shown in the SM for several values of activity~\cite{SM}. The critical point corresponds to $\Delta\phi(a_c^*,\phi_0^c)=0$ and decreases with increasing activity strength, as evident from Fig.~\ref{fig:pl2d} (see also inset of Fig.~(3) in the SI). Figure~\ref{fig:fss} shows a finite size scaling of $\Delta\phi(a,\phi_0^c)$ with system size that suggests a continuous phase transition with a diverging $d\Delta\phi/da$ at the critical point. A system size of $L=128$ approximates well the asymptotic behavior.

\emph{Interfacial dynamics and linear stability.} 
We show that simple scaling arguments can be used to obtain the growth laws of domains in the coexistence region.
To estimate the growth rate of droplets starting from a uniform state, we assume that growth is controlled by a single length scale $L$, moving at a velocity $v$, so that $\dot L\sim v$. If the dynamics of the system is dominated by diffusive currents, we estimate $v\sim \nabla\mu\sim M\sigma/L^2$, where $\sigma=\left(-8\kappa a^3/(9b^2)\right)^{\red{1/2}}$ is the interfacial tension \cite{kendon2001inertial}. If the dynamics is dominated by flow, then $|v|$ can be estimated from the Stokes equation by balancing viscous stress with elastic and active stresses, as $\eta v/L^2\sim \sigma_a/L^2+\alpha /L + K\lambda/L^3$, \blue{where, for convenience, we have defined an effective active interface tension $\sigma_a=\kappa_a\sigma/\kappa$. This active tension arises from interfacial active stresses and can be negative when $\kappa_a<0$, resulting in a self-shearing instability discussed previously in the literature~\cite{Singh2019}.} We can then write the rate of change of $L$ as
\eq
\label{eq:dotL}
\dot L = s_1\frac{M\sigma}{L^2}+s_2\frac{\sigma_a}{\eta}+s_3\frac{\lambda K}{\eta L}+s_4\frac{\alpha L}{\eta},
\eeq
where $s_i$ are unknown constants. For scalar models with no coupling to nematic degrees of freedom, only the first two terms appear in Eq.~\eqref{eq:dotL}. Active interfacial stresses with $\sigma_a<0$ can then arrest coarsening through \rev{a self-shearing instability}, as discussed in \cite{Singh2019}. This results in domains of typical size $L^*\sim (M\sigma/|\sigma_a|)^{1/2}$. When, however, the active fluid is an extensile active nematic, bulk active stresses of strength $\alpha<0$ lead to a different path to arrested phase separation. Assuming for simplicity $\sigma_a=\sigma$, the first three terms on the right hand side of Eq.~\eqref{eq:dotL} are positive, hence describe coarsening, while the last is negative for extensile activity and arrests domain growth.

We can identify three regimes:\\
(i) In the early stages of coarsening growth is controlled by diffusion of material from smaller to bigger droplets, with $L(t)\sim t^{1/3}$. If activity is large enough,  growth is arrested at short times when active currents, of order $j_a\sim \alpha L/\eta$, balance diffusive currents of order $j_d\sim M\sigma/L^2$, resulting in $L^*\sim\ell_D= (M\sigma\eta/|\alpha|)^{1/3}$.\\
(ii) At intermediate times elastic nematic stresses transmit interactions that drive domain growth, with $L(t)\sim t^{1/2}$. This mechanism becomes relevant for large nematic stiffness $K$, resulting in droplets size controlled by the active length $L^*\sim\ell_a=(K/|\alpha|)^{1/2}$.\\
(iii) Finally, at late times drops coalesce through hydrodynamic advection and $L(t)\sim t$. In this regime growth is arrested when interfacial tension balances active stress, which gives $L^*\sim\ell_\sigma=\sigma/|\alpha|$.

Whether these regimes are all accessible experimentally depends of course on parameter values. A detailed study of the coarsening dynamics is left for future work. An important prediction, however, is the dependence of the steady state droplet size on activity $\alpha$, which may  provide  a  criterion for sorting out the most relevant active mechanisms in experimental realizations.

The scaling arguments presented above are supported by an analysis of the linear stability of \red{a} flat interface at $y=0$. The interfacial height, $h(x,t)$, is defined by writing the phase field as $\phi(r,t) = f(y+h(x,t))$, such that $f(u)$ changes sharply at $u=0$, i.e. $f'(u)\approx\delta(u)$. We assume that the active fluid occupying the region $y>0$ is in an initial uniform nematic state aligned with the interface, justified because extensile active nematics tend to align with interfaces through a process called active anchoring \cite{Blow2014,Coelho2021}. We can then write $Q_{xx}\approx S(u)/2$ and $Q_{xy}\approx S(u)\theta(x,t)$, where $S(u)$ is the amplitude of the nematic order parameter, with $S'(u)\approx\delta(u)$, and $\theta(x,t)$ the small angle of the nematic director with respect to the interface. We eliminate the velocity $v_i$ by assuming low Reynolds number and solving for the Stokes flow by imposing incompressibility. The Fourier components of the velocity are 
$v_i(\mathbf{q},t) = P_{ij}(\mathbf{q})f_j(q,t)/\eta q^2$,
with $P_{ij}(\mathbf{q})=\delta_{ij}-\hat{q}_i\hat{q}_j$ a projection operator and $\hat{q}_i=q_i/|\mathbf{q}|$. Following \cite{Bray2002,Fausti2021}, we can then obtain linear equations for $h(q_x,t)$ and $\theta(q_x,t)$, given by \cite{SM}
\eq
\partial_t \binom{h}{\theta} = A\binom{h}{\theta},
\eeq
with
\eq\label{eq:matrixA}
A = \left[ \begin{array}{cc}
    -\frac{\sigma_a|q_x|}{4\eta} - \frac{M\sigma|q_x|^3}{2} & -\frac{\alpha iq_x}{2\eta q_x^2} \\
    \frac{\kappa_a\sigma|q_x|iq_x}{4\kappa\eta}+\frac{\alpha(1-\lambda)iq_x}{8\eta} & -\frac{\alpha\lambda}{2\eta}-\frac{K}{\gamma}q_x^2 \\
  \end{array} \right]\;. 
\eeq
For $\alpha=0$, height and director fluctuations are decoupled. Height fluctuations become unstable for $\sigma_a<0$ below a wavenumber $q^*\sim\sqrt{|\sigma_a|/(\sigma\eta M)}$, signaling
arrested phase separation with structures of size $L^*\sim 2\pi/q^*$. 
Bulk extensile activity ($\alpha<0$) destabilizes director fluctuations on all scales according to the generic instability of bulk active nematics~\cite{PhysRevLett.89.058101}. When $\sigma_a>0$, the coupling to interfacial relaxation can stabilize the nematic above a characteristic length scale, often in a regime where the modes become complex, signaling the propagation or surface capillary-like waves \cite{soni2019stability,ray_unpublished}. 

\begin{figure}
    \includegraphics[scale=0.2]{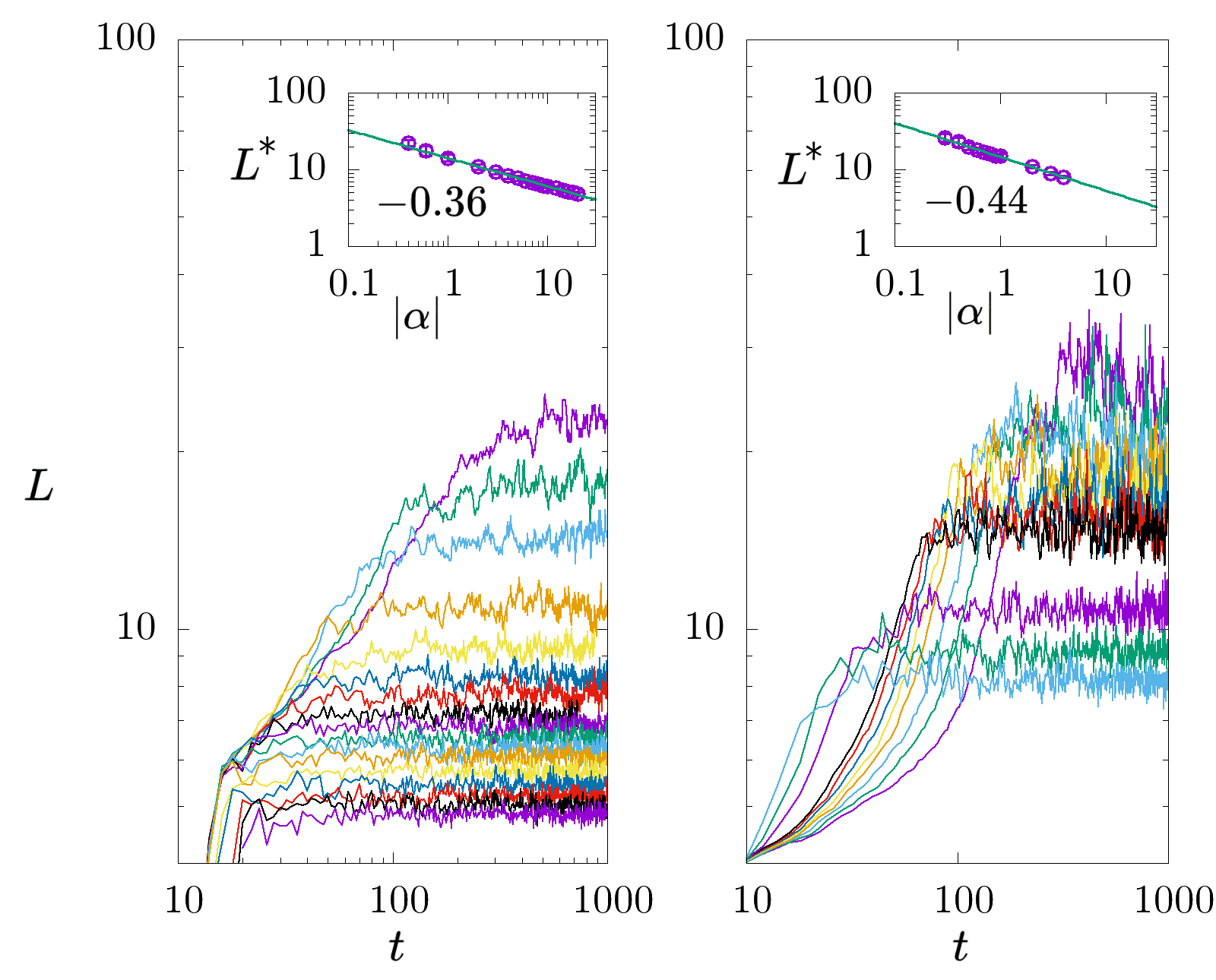}
    \caption{\label{fig:powerlaws}\blue{Droplet size as a function of time for an initially uniform state and various activity for two sets of parameters. Left: $l_D\approx 10 l_{\alpha}$ (left), i.e., regime (i) where coarsening is controlled by diffusion and we estimate $L^*\sim |\alpha|^{-1/3}$. Right: $l_{\alpha}\approx 10 l_D$ (right), i.e., regime (ii) where coarsening is controlled by elasticity and we estimate $L^*\sim |\alpha|^{-1/2}$. The insets show the scaling  of the saturation value $L^*$ with $|\alpha|$.
    }}
\end{figure}

A  detailed analysis is presented in the SI. As expected on the basis of scaling arguments, $q^*$ is controlled by the three active length scales (for  $\sigma_a=\sigma$) $\ell_D,\ell_a,\ell_\sigma$, with crossovers between the three scalings with increasing activity. \blue{Figure \ref{fig:powerlaws} shows the growth in time of droplet size in the micro-phase separated state for various activities, and the scaling of $L^*$ with activity (inset) for two set of parameters corresponding to regimes (i) and (ii).}

In summary, we have shown that interfacial and bulk active stresses play distinct roles in the LLPS of active/passive mixtures. Previously studied interfacial stresses can arrest the phase separation via a shearing instability of the interface, but do not affect the location of the critical point. This is because their action is confined at the interface, where the concentration gradients are finite. In contrast, bulk extensile active stresses can suppress phase separation entirely by  sustained self-stirring of the active component.  As a result, bulk active stresses can significantly shift the critical point of the passive phase separating mixture to lower temperatures and shrink the size of the coexistence region in the volume fraction/temperature plane.
Within the coexistence region,  phase separation is arrested leading to a emulsion-like steady state  of dynamic droplets that continuously coalesce and split due to the shear flows created by the active nematic. 

Our work quantifies the role of activity on the phase diagram of phase separating active-passive binary mixtures. It highlights the distinct roles of interfacial and bulk active stresses and it shows that bulk active stresses coupling to liquid crystalline degrees of freedom can \red{shift the critical point.}
Our results on the scaling of the typical droplet size as a function of activity in the dynamical emulsion-like state offer quantitative criteria for discerning the mechanisms that arrest phase separation in experiments and simulations.

\begin{acknowledgments}
{\em Acknowledgments:}
We thank Zhihong You, Cesare Nardini, Alexandra Tayar and Zvonimir Dogic for illuminating discussions. This work was supported by NSF grant DMR-2041459.
\end{acknowledgments}

\bibliography{refs}

\end{document}


\title{Supplementary information for Activity suppressed phase separation}

\author{Fernando Caballero}
\email{fmc36@ucsb.edu}
\affiliation{Department of Physics, University of California Santa Barbara, Santa Barbara, CA 93106}
\author{M. Cristina Marchetti}
\affiliation{Department of Physics, University of California Santa Barbara, Santa Barbara, CA 93106}
\date{\today}
\maketitle
    \renewcommand{\thefigure}{S\arabic{figure}}

\setcounter{figure}{0}
    \section{Extensile and contractile stresses in an active nematic and active model H}

\begin{center}
        \begin{figure}[b]
            \includegraphics[scale=1]{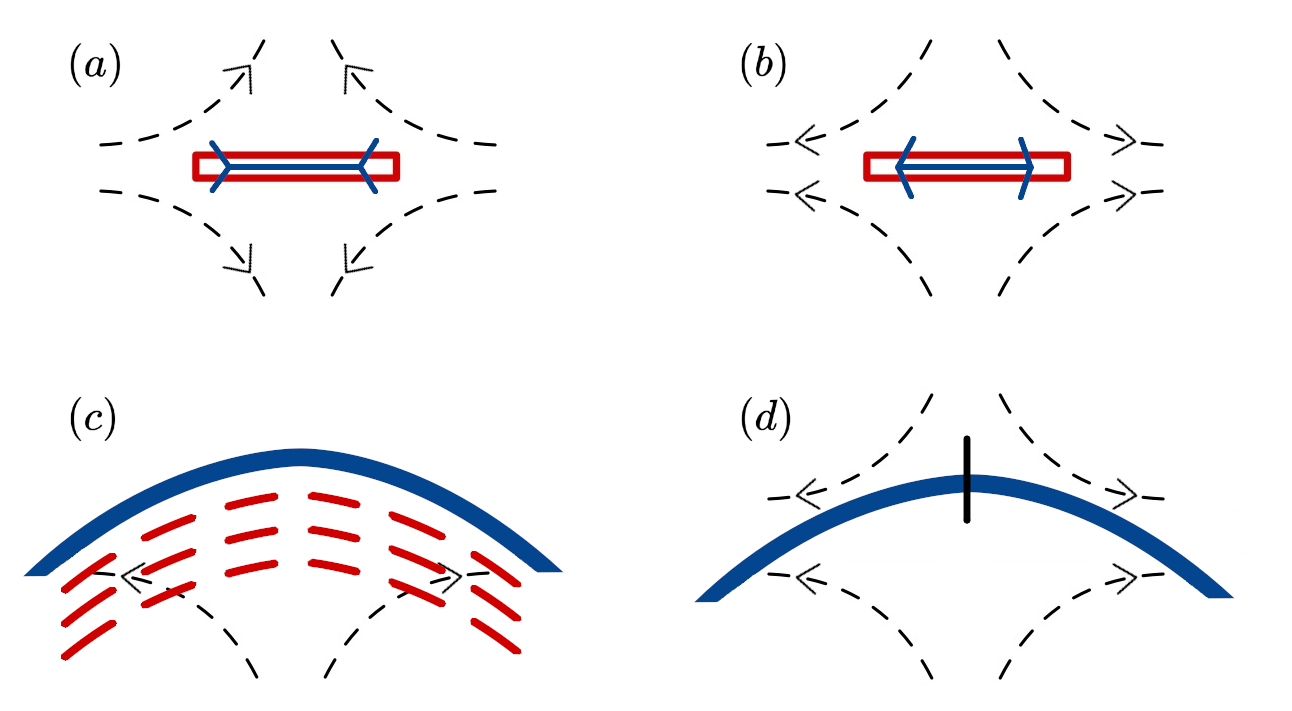}
            \caption{\label{fig:sisketches}These sketches represent the different stresses we analyse. In $(a)$ and $(b)$ we see the forces on the fluid around a rod (in red) for contractile and extensile active stresses respectively, with the force dipole depicted in blue. In $(c)$ we sketch the effect of a nematic (in red rods) aligned with a curved interface (in blue) for an extensile stress, while in $(d)$ we see the force on the flow by a stress of the form of $Q^\phi$ in active model H, where its director $(\nabla\phi)$ is marked by the black rod perpendicular to the interface, the forces are contractile for the black rod, while extensile with respect to the interface, as explained in the text.}
        \end{figure}
    \end{center}

    This section includes a short summary of extensile and contractile stresses for both models for the reader's convenience. The main text focuses mainly on the bulk active stress created by the liquid-crystalline degrees of freedom, although we also briefly mention and discuss a purely interfacial stress. The bulk active stress of an active nematic is written as $\sigma_Q=\alpha Q_{ij}$, where positive and negative values of $\alpha$ describe contractile and extensile stresses respectively. If the constituent particles making the active nematic are uniaxial rods, these stresses are of the form sketched in Figure \ref{fig:sisketches} (a) and (b). We have additionally focused only on the extensile case, as uniform states are generally unstable to these stresses~\cite{PhysRevLett.89.058101} and develop chaotic flows that provide the self-stirring force that stabilizes the uniform state.
    
    The second form of activity is that of active model H, produced by allowing the stress of interface gradients to have non equilibrium values. In the main text we include this activity through the parameter $\sigma_a=\kappa_a\sigma/\kappa$, which has its equilibrium value at $\kappa_a=\kappa$. Notice that the form of the stress, $\sigma_\phi = -\kappa_a(\partial_i\phi\partial_j\phi -\delta_{ij}/2(\nabla\phi)^2)$, resembles the bulk stress for a nematic director in the direction of $\partial_i\phi$. Physically, we can then interpret negative values of $\kappa_a$ as contractile stresses in the direction of $\partial_i\phi$, or extensile in a direction tangential to the interface, and viceversa for positive values of $\kappa_a$. This is sketched in Figure \ref{fig:sisketches} (c) and (d). In the case of active model H, however, it is self-shearing of the interface that drives the dynamics of the system, while in the case of an active liquid crystal is shearing of the flow also in the bulk region by the liquid crystal that drives the dynamics. This difference turns out to be crucial in shifting the critical point, as explained in the main text and below.
    
    Notice that active model H also generally includes other terms that function as time reversal symmetry breaking terms in the chemical potential of $\phi$~\cite{Tiribocchi2015}. We have ignored these terms as they do not have an impact in the flows created by interfacial stresses nor in the typical growth laws observed~\cite{Tiribocchi2015,Singh2019}. 

    \section{Interface activity effect on $\Delta\phi$}
    
    In this section we discuss the effect of $\kappa_a$ on the shape of the phase diagram. In the main text, we mention this interfacial activity is not enough to shift the critical point. In a purely scalar model, like active model H, activity can only come as stresses on the flow that must necessarily be a function of $\nabla\phi$, or terms in the diffusive current of $\phi$ that explicitly break TRS. Non gradient terms cannot form part of the latter contribution since these can always be written as the derivative of a free energy and therefore cannot break TRS. This means that the strength of any such term decreases as $\Delta\phi$ decreases, and completely disappear for a fully uniform state. Uniform states are therefore always linearly unstable to density fluctuations and we should not expect the critical point to be shifted. We can, however, expect the coexistence densities to change, since there could be an interplay between diffusive phase separating dynamics and the active stresses created by density gradients that balance each other at some finite $\Delta\phi$. 
    
    For this reason, in the main text we suggest that an additional degree of freedom is necessary to stabilize the uniform state, in this case an active nematic that remains strongly non-uniform in the steady state, providing a self-stirring force to the system.
    
\begin{figure}
        \centering
        \includegraphics[scale=0.2]{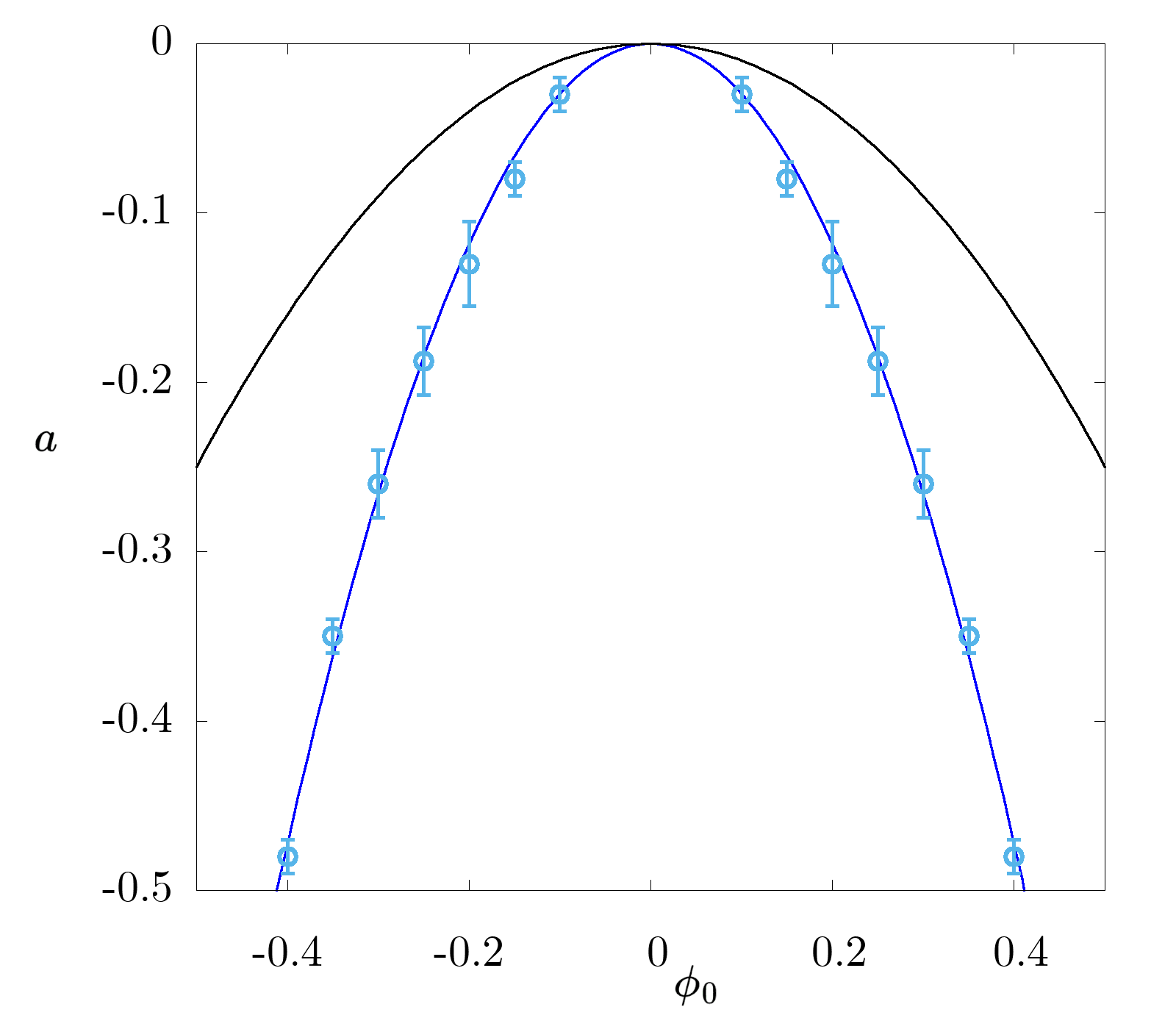}
        \caption{Effect of interfacial activity on the phase diagram in active model H ($\alpha=0$ and $\sigma'<0$ in our model). The black line is the mean field coexistence region of the equilibrium model, while the points are the results of a numerical scan, equivalent to that found in Figure 1 of the main text. Notice that the coexistence densities are indeed changed by activity, but the critical point is not. Only the points for $\phi_0>0$ are calculated and then mirrored, since model H is symmetric in $\phi\rightarrow-\phi$.}
        \label{fig:si_interface}
    \end{figure}

    Figure \ref{fig:si_interface} shows the effect of interface activity on active model H, showing that the critical point is not shifted, but the value of $\Delta\phi$ below the shifted critical point does indeed change. The simulations of Fig~\ref{fig:si_interface} have been done starting at a full phase separated state, and letting it evolve with $\kappa_a<0$. The parabolic fit to the data is very close to the spinodal line, suggesting the effect of activity in the phase diagram of active model H is turning the metastable phase separated states of the equilibrium model H into unstable ones.
    
    \section{Effect of increasing activity on the critical point}
    This section gives a few more details on the behaviour of the critical point as a function of activity. As mentioned in the main text, and shown in Fig 1 of the main text, increasing activity increases the magnitude of the downward shift of the critical point, while also shifting it to lower values of volume fraction (or higher relative density of the active species). A precise characterization of this critical point, both in its position and critical properties, is a hard numerical problem due to the long range and time correlations, therefore in this Letter we only approximated its position by running numerical simulations in the arrested phase separated state, and then increasing temperature $a$ until we observe, visually, a uniform state. This is usually clear far from the critical point due to the discontinuous nature of the transition, but hard close to the critical point where the transition becomes continuous.
    
\begin{figure}
        \includegraphics[scale=0.18]{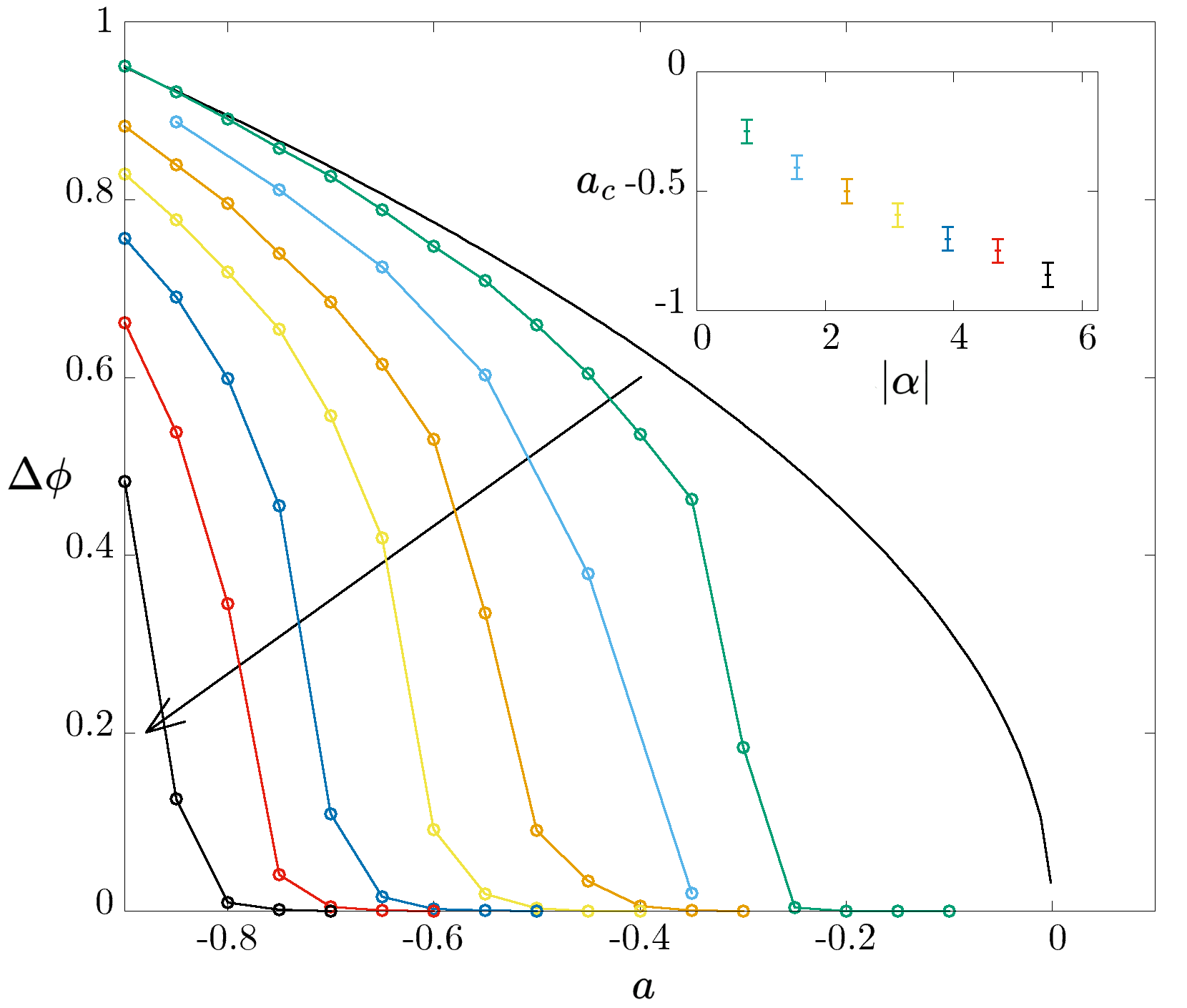}
        \caption{\label{fig:act_crit} Late time values of $\Delta\phi$ as a function of $a$ for increasing values of $|\alpha|$, as indicated by  the arrow, with values going from  $|\alpha|=0.8$ to $|\alpha|=5.5$. The black line is the mean field equilibrium result, given by $\Delta\phi=\sqrt{-a}$. The inset shows the critical point $a_c$ as a function of activity. The colors in the inset correspond to colors in the main plot. }
    \end{figure}

    The way to locate the critical point, for different values of activity, would be then to do an approximate scan in the temperature/relative concentration plane, as in Fig 1 of the main text. With that data we can use a parabolic fit $a(\phi_0)$ to find the approximate critical point as the maximum point $\phi_0^c$, at which $\left.da(\phi_0)/d\phi_0\right|_{\phi_0^c} = 0$. Now we run simulations scanning for different values of $a$ at that volume fraction $\phi_0=\phi_0^c$, and look at how $\Delta\phi$ depends on $a$. Doing so gives data as shown in Fig \ref{fig:act_crit}, where $\Delta\phi(a)$ curves are plotted for several values of activity, showing the very clear increase in the shift of the critical point as we increase $|\alpha$. The critical point is the point $a_c$ at which $\Delta\phi(a_c)=0$, as we increase $|\alpha|$.

    \section{Linear interface dynamics for active model H}
    This section shows the linearization of the interface dynamics of active model H for an initially flat interface. This result helps build the linear dynamics of an initially flat interface in the full model with liquid-crystalline degrees of freedom (Eqs (5-6) in the main text). This computation in the equilibrium case is done in~\cite{Bray2002}, and is mathematically equivalent to the case presented in this section, except for a constrain in the parameters $\kappa$ and $\kappa_a$, since equilibrium requires $\kappa = \kappa_a$. Additionally, we consider the case for finite substrate friction $\Gamma$, although we set it to $0$ in the analysis of the main text.

    The equations of motion are
    \aeq
    \partial_t\phi + v\cdot \nabla \phi &= M\nabla^2\left(a\phi+b\phi^3-\kappa\nabla^2\phi\right),\\
    0 &= -\Gamma v + \eta\nabla^2v - \nabla P +\mu\nabla\phi,
    \eaeq
    where we have assumed very low Reynolds number to approximate the velocity to that of the Stokes solution to the NS equation. Notice that the force $\mu\nabla\phi$ can be written as the divergence of a stress $\nabla\cdot\sigma$, with $\sigma_{ij}=-\kappa_a(\partial_i\phi\partial_j\phi-\delta_{ij}/d(\nabla\phi)^2)$ by absorbing constant terms in the pressure~\cite{Cates2018}.

    Assuming an initially flat interface with height function $h(x,t)$, we first define the function $f(u) = \phi(y+h(x,t))$, where we have chosen $y$ to be the coordinate perpendicular to the interface. We first calculate the contribution of advection to the linear dynamics of $h$. The first step is to formally solve the Stokes equation, for which the velocity can be written as
    \eq
    v_i = \int d^dr' T_{ij}(r-r')(\partial_j\phi(r'))\mu(r'),
    \eeq
    where $T$ is the Oseen tensor, which has the following Fourier form in our case
    \eq
        T_{ij} = \frac{1}{\Gamma+q^2\eta}\left(\delta_{ij}-\frac{q_iq_j}{q^2}\right)
    \eeq
    Substituting $\phi$ for $f$, and noting that
    \aeq
    \partial_t\phi &= f'(u)\partial_th, \\
    \partial_i\phi &= f'(u)\left(\textbf{e}_y + \nabla_xh\textbf{e}_x\right) \\
    \kappa_a\nabla^2\phi & = \kappa_a(-f'(u)\nabla_x^2h+f''(1+(\nabla_xh)^2)),
    \eaeq
    we can then write the linear part of the dynamics corresponding to the advection term, $h^{(\text{adv}}$ as follows
    \eq
        f'\partial_t h^{(\text{adv})}(x,t) = \kappa\int d^dr' f'(u)T_{yy}(r-r')f'(v)\nabla_{x'}^2h(x',t),
    \eeq
    where $u = y+h(x,t)$ and $v=y'+h(x',t)$. The only term to linear order is the term involving $T_{yy}$. To move forward, we assume the interface to be thin, so that we can approximate the function $f'$ to be, initially, a Dirac $\delta$ function in $y$. We also now transform our equation in Fourier space in the $x$ plane, which can be done by using the Fourier transform of $T_{ij}$ and transforming back to Real space only in the coordinate $y$. Lastly, we get the linear evolution equation for $h$ by setting $y=0$, or alternatively, since $f'$ is not well defined at $0$, by multiplying the equation with $f'(u)$, and integrating along $u$, and using the definition of the interface function as the free energy cost of creating interfaces $\sigma = \kappa\int_{-\infty}^\infty [f'(u)]^2du$. The final result is
    \eq
        \partial_t h^{(\text{adv})}(q_x,t) = -\frac{\kappa_a \sigma}{2\kappa}\frac{|q_x|^3}{\Gamma+\eta|q_x|^2+\sqrt{\eta}|q_x|\sqrt{\Gamma+\eta |q_x|^2}}h^{(\text{adv})}(q_x,t).
    \eeq

    Notice this result reduces to the known equilibrium results of~\cite{Bray2002} by setting $\kappa=\kappa_a$, and looking at the limits $\Gamma\rightarrow 0$ and $\eta\rightarrow 0$. The first limit gives the behaviour of model H, with interface fluctuations decaying as $\dot h\sim -qh$, and gives the hydrodynamic domain growth rate $L\sim t$. The second limit, equivalent to a diffusive behaviour of $\phi$, since there is no viscosity, gives the dynamics of model B, with fluctuations decaying with the cube of $q$; $\dot h \sim -q^3h$, and gives the domain growth rate of model B $L\sim t^{1/3}$.
    The contribution to $\dot h$ given by the diffuse dynamics of $\phi$ is calculated in~\cite{Bray2002}, and gives a contribution
    \eq
    \partial_t h^{(\text{dif})}(q_x,t) = -\frac{M\sigma|q|^3}{2}h(q_x,t)
    \eeq

    Setting $\Gamma=0$ and adding both contributions gives the first diagonal term of matrix $A$ of the main text responsible for the relaxation of $h(q,t)$. Notice that if we set $\Gamma=0$, the full dynamics of $h$ becomes
    \eq
    \partial_t h(q,t) =  -\frac{\sigma_a}{4\eta}|q|h(q,t) - \frac{M\sigma}{2}|q|^3h(q,t). 
    \eeq
    
    This expression provides the growth laws corresponding to hydrodynamic effects, if for instance $M=0$, ($L\sim t$), and to diffusive effects, if $\sigma_a=0$ ($L\sim t^{1/3}$). In the active extensile case, i.e. $\sigma_a<0$, modes for $q<q_c$ are unstable, with $q_c = (|\sigma_a|/(\eta M\sigma))^{1/2}$, which gives the steady state mean droplet size for active model H derived in~\cite{Singh2019} through dimensional analysis, $L\sim |\sigma_a|^{-1/2}$.
    
    \section{Full linear interface dynamics}

    The equations for both the height function $h$ and the nematic director angle $\theta$, equations (5-6) in the main text, are obtained using the same method as in last section.

    One linearizes the equation for the velocity to obtain, $v_i = \int d^dr' T_{ij}(r-r')f_j(r')$, where the force now has two terms, $f_j = -\kappa_a \partial_k(\partial_j\phi\partial_k\phi-\delta_{jk}/d(\nabla\phi)^2) + \partial_k( \alpha\tilde\phi Q_{jk})$. The first term is the one of active model H, responsible for interface stresses produced by density gradients, and the second is the bulk active stress produced by the active nematic. Notice that $\tilde\phi$ in the bulk stress will not contribute to linear dynamics, since it only acts as a step function at early times, and will produce only higher order nonlinearities of order $O(Q\partial h)$ and $O(h\partial Q)$, so we can ignore its effect on the bulk stress in the linear expansion we calculate below.

    Since we assumed the nematic is initially aligned with the interface, $\theta$ is small, and we can approximate $Q_{xx}\approx S/2+O(\theta^2)$, and $Q_{xy}\approx S\theta+O(\theta^2)$

    We can now write out the linear part of the velocity
    \aeq 
    v_i(r) =& - \frac{\kappa_a\sigma}{\kappa}\int dx' T_{iy}(x-x', y)\nabla_{\xb'}^2h(\xb',t) \\
            & + \alpha\int dr' T_{ij}(r-r')\partial_kQ_{jk}(r')
    \eaeq

    This linear part is what we use to calculate the various contributions to the behaviour of the height and nematic close to the interface. First we calculate the effect of advection on $h$, which comes from the equation for $\phi$,
    \eq
    f'(u)\partial_th(x,t) = - v_i\partial_i\phi = \frac{\kappa_a\sigma}{\kappa}\int dx' f'(u) T_{yy}(x-x',y)\nabla_{\xb'}^2h(\xb,t) - \alpha\int dr' f'(u) T_{yj}(r-r')\partial_kQ_{yk}(r')
    \eeq

    To find the dynamics at the interface we multiply with $f'(u)$ and integrate accross the interface. Remember integrals of the delta function f'(u) are of strength $2$ ($\phi$ jumps from $-1$ to $1$).

\eq
\partial_t h(x,t) = \frac{\kappa_a\sigma}{\kappa}\int dx'T_{yy}(x-x',0)\nabla_{x'}^2h(x',t) -\alpha\int dr' T_{yj}(x-x',-y')\partial_kQ_{yk}(r').
\eeq

It's useful to compute each term separately. The first one is easy, we can use that $T_{yy}$ in Fourier space has the form $(2\pi)^{-d}(\eta q^2)^{-1}(1- q_y^2/q^2)$ to get to 

\eq
\sigma\int dx'T_{yy}(x-x',0)\nabla_{x'}^2h(x,t) = -\frac{\kappa_a\sigma}{4\kappa\eta}\int dq_x e^{i q_x x} |q_x| h(q_x,t),
\eeq

where we explicitly computed the integral of the Oseen tensor in 1 dimension

\eq
\left.\int dq_y e^{i q_y y}\frac{1}{(2\pi)^d}\frac{1}{\eta q^2}\left(1-\frac{q_y^2}{q^2}\right)\right|_{y\rightarrow 0} = \left.\frac{1}{(2\pi)^{d-1}4\eta}e^{-|y||q_x|} \left(| y| +\frac{1}{|q_x|}\right)\right|_{y\rightarrow 0} = \frac{1}{(2\pi)^{d-1}}\frac{1}{4\eta}\frac{1}{|q_x|}
\eeq
The second term is more complicated

\aeq
-\alpha\int dr' T_{yj}(x-x',-y')\partial_kQ_{yk}(r') = & -\alpha\int dr' T_{yx}(x-x',-y')\left(\partial_xQ_{xx} + \partial_yQ_{xy}\right)\\
  & -\alpha\int dr' T_{yy}(x-x',-y')\left(\partial_xQ_{xy}+\partial_yQ_{yy}\right)
\eaeq

Expanding the derivatives to linear order, we get
\aeq
-\alpha\int dr' T_{yj}(x-x',-y')\partial_kQ_{yk}(r') = & -\alpha\int dr' T_{yx}(x-x',-y')s'(u')\left(\frac{1}{2}\partial_{x'}h(x',t) + \theta(x',t)\right)\\
 & -\alpha\int dr' T_{yy}(x-x',-y')s(u')\partial_{x'}\theta(x',t) 
\eaeq
We have ignored $\partial_yQ_yy\propto s'(u')$, since it gives a term in the velocity proportional to $\delta(0)$ that thus vanishes for any finite wavelength. (This needs careful checks).

Notice the first integral has a $s'(u')$ factor, which is now a delta funciton of strength one, allowing to calculate the integral in $dy'$ directly. Again we need the integral of the Oseen tensor in 1D, now of its $T_{xy}$ component:
\eq
\int dq_y e^{i q_y y}\frac{1}{(2\pi)^d}\frac{1}{\eta q^2}\left(1-\frac{q_y^2}{q^2}\right) = \frac{1}{(2\pi)^{d-1}}\frac{1}{4\eta}e^{-|q_x||y|}\frac{-i q_x y}{|q_x|}
\eeq

This integral is proportional to $y$, so it will vanishes when integrated with $f'(u)$. The second one has to be computed with the integral of $T_{yy}$, written above. The final integral involving $y'$ is $\int_0^\infty dy' \exp(-|q_x| y')(|q_x|^{-1}+y) = 2|q_x|^{-2}$. The final equation for $h(q_x,t)$ is then                                                                 
                                                                                              
\eq                                                                                           
\partial_th^{\text{(adv)}}(q_x,t) = - \frac{\sigma_a}{4\eta}|q_x|h(q_x,t) -\frac{\alpha}{2\eta}\frac{iq_x}{|q_x|^2}\theta(q_x,t),
\eeq
as with model H, the effect of the diffusive dynamics of $\phi$ provides a term $\partial_th^{\text{(dif)}}=-\frac{M\sigma}{2} |q|^3$. In this case there is no $\kappa_a$ term since this is only present in the active stress in the Navier-Stokes equation, and not in the chemical potential.

The calculation is similar for the angle $\theta(q,t)$. Since the bulk nematic is ordered, the relaxational dynamics is purely elastic, giving the term $-Kq^2$ of the diagonal of the matrix $A$. The rest of the terms are given by the coupling of $Q$ to the flow $v$, through $\lambda D_{ij}$ and $[Q,\omega]_{ij}$. The dynamics of $\theta$ to linear order can be written as

\eq
    S(u)\partial_t\theta(x,t) = \frac{\lambda}{2}(\partial_xv_y+\partial_yv_x)+\frac{S(u)}{2}(\partial_xv_y-\partial_yv_x) +K\nabla^2\theta(x,t)
\eeq

To eliminate $S(u)$, as we did with $h(x,t)$, we multiply the equation with $S'(u-u_0)$, where $u_0$ is a distance from the interface large enough that the nematic is already developed, so that $S(u_0)=1$, but very close to the interface. Since we approximate $S(u)$ to be a step function, as we did with $f(u)$, $u_0$ can be taken to be very small.  Integrating accross the interface will then eliminate all factors of $S$, and the equation of $\theta(q,t)$ in Fourier space on the interface plane is 

\eq
\partial_t\theta(q,t) = \frac{\lambda}{2}\left(iq_xv_y+\partial_yv_x\right)+\frac{1}{2}\left(iq_xv_y-\partial_yv_x\right) - Kq^2\theta(q,t)
\eeq

The first component of the velocity, $v_x$, can be lienarly expanded in $h$ and $\theta$ as follows

\aeq
v_x = & - \frac{\kappa_a\sigma}{\kappa} \int dx' T_{xy}(x-x',y)\nabla_{x'}^2h(x',t) \\
      & + \alpha \int dr' T_{xj}(r-r')\partial_kQ_{jk}(r').
\eaeq

The term in the first line, which corresponds to tangential velocities created by interface gradients, is
\eq
- \frac{\kappa_a\sigma}{\kappa} \int dx' T_{xy}(x-x',y)\nabla_{x'}^2h(x',t)=-\frac{\kappa_a\sigma}{4\kappa\eta}\int dq_x e^{iq_x x}|q_x|(iq_x)y e^{-|q_x||y|} h(q_x,t)
\eeq

The second term, responsible for the effect of bulk active stresses on the angle close to the interface is
\aeq
\alpha \int dr' T_{xj}(r-r')\partial_kQ_{jk}(r') &= \alpha \int dx' T_{xx}(x-x',y)\left(\frac{1}{2}\partial_{x'}h(x',t) + \theta(x',t)\right) \\
& + \alpha\int dr' T_{xy}(r-r')s(u')\partial_{x'}\theta(x',t) \\
& = \frac{\alpha}{4\eta}\int dq_x e^{iq_x x} e^{-|q_x||y|}\left(\frac{1}{|q_x|}-|y|\right)\left(\frac{1}{2}(iq_x)h(q_x,t)+\theta(q_x,t)\right)\\
& + \frac{\alpha}{4\eta}\int dy'dq_x e^{iq_xx}e^{-|q_x||y-y'|}|q_x|(y-y')s(u')\theta(q_x,t)
\eaeq

Finally, $v_y$ is 
\aeq
v_y = & - \frac{\kappa_a\sigma}{\kappa}\int dx' T_{yy}(x-x',y)\nabla_{x'}^2h(x',t) \\
& +\alpha\int dr' T_{yj}(r-r')\partial_kQ_{jk}(r')
\eaeq
for which both terms can be calculated as for $v_x$, to give
\aeq
v_y = & \frac{\kappa_a\sigma}{4\kappa\eta} \int dq_x e^{iq_x}e^{-|q_x||y|}\left(\frac{1}{|q_x|}+|y|\right)|q_x|^2h(q_x,t) \\
& + \frac{\alpha}{4\eta}\int dq_xe^{iq_xx}e^{-|q_x||y|}\frac{(-iq_x)y}{|q_x|}\left(\frac{1}{2}(iq_x)h(q_x,t)+\theta(q_x,t)\right) \\
& + \frac{\alpha}{4\eta}\int dy'dq_xe^{iq_xx}e^{-|q_x||y-y'|}\left(\frac{1}{|q_x|}+|y-y'|\right)s(u')(iq_x)\theta(q_x,t).
\eaeq

Putting these expressions for $v_x$ and $v_y$ in the equation of $\partial_t\theta(q,t)$, we get
\eq
\partial_t\theta(q,t) = \left(\frac{\sigma_a|q|}{4\eta}+\frac{\alpha(1-\lambda)}{8\eta}\right)iqh(q,t) -\frac{\alpha\lambda}{2\eta}\theta(q,t) - Kq^2\theta(q,t)
\eeq

Equations (23) and (31) give matrix $A$ in equation (6) of the main text.

\section{Analysis of the linear dynamics of the interface}

\begin{figure}
    \includegraphics[scale=0.2]{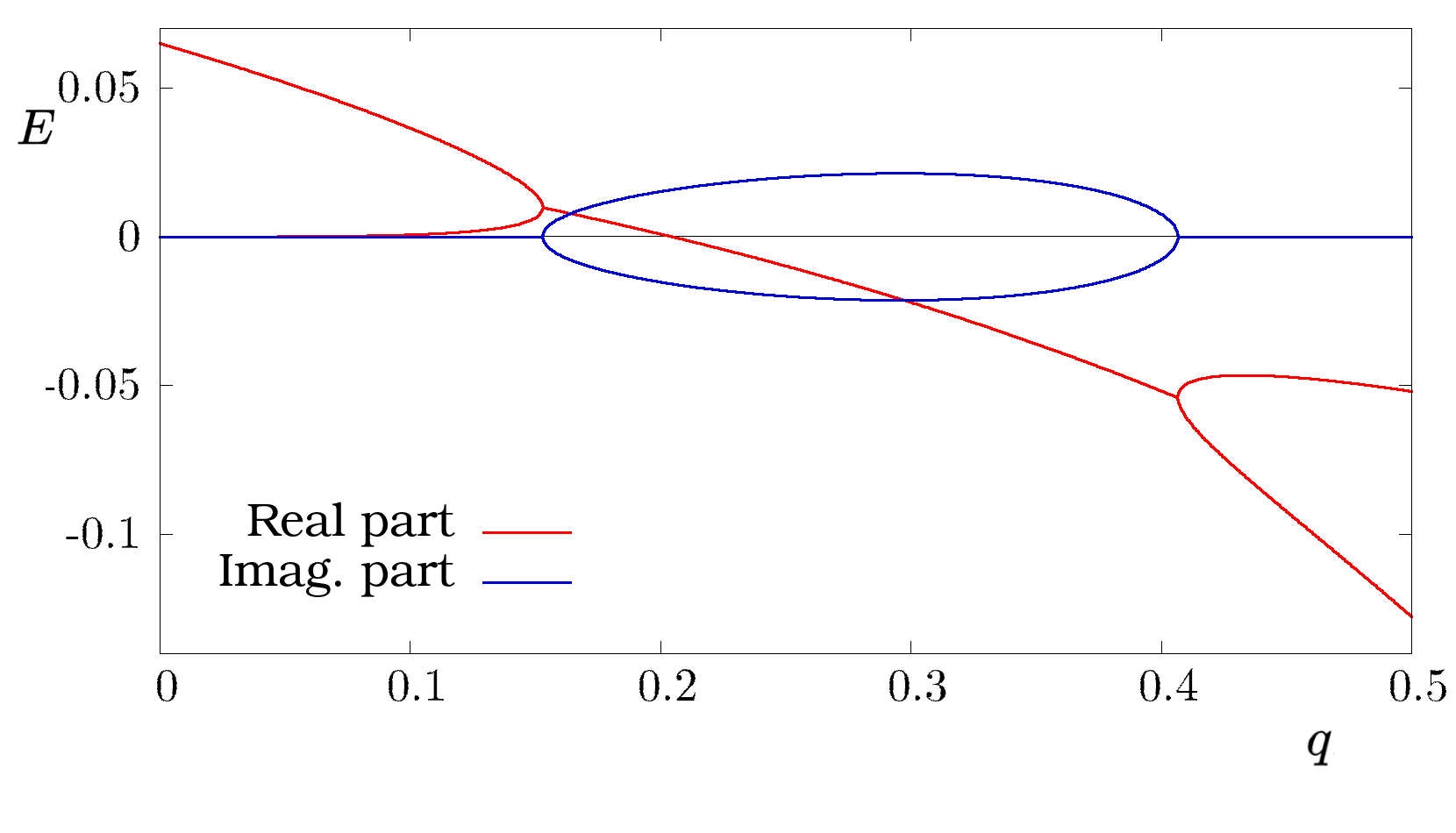}
    \caption{\label{fig:si_disp} Real(red) and imaginary (blue) part of the}  eigenvalues $E$ of the dynamical matrix $A$ given in Eq.~\eqref{eq:matrixA} for  $\kappa_a=\kappa=\sigma=M=\lambda=1$, $\alpha=-0.1$, $K=0.2$. 
\end{figure}

\begin{figure}
    \includegraphics[scale=0.2]{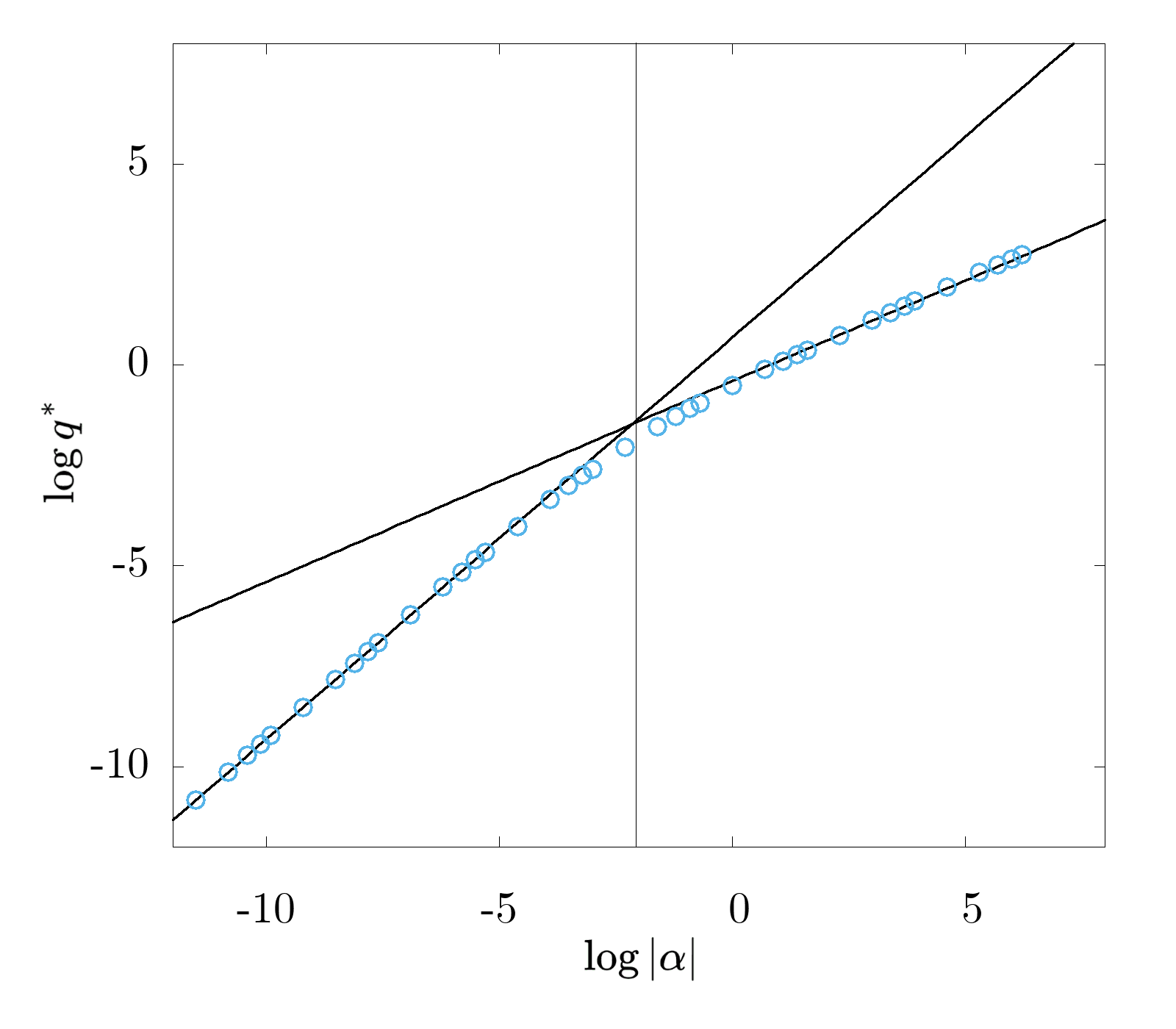}
    \caption{\label{fig:qstar} Value of the wavevector at which the real part of the higher eigenvalue changes sign, thus describing the expected size of droplets as a function of the activity parameter $\alpha$. This has been obtained by numerically solving the equation $E_+(q)=0$, with $M=0$ to ignore diffusive effects, and $\sigma=K=1$, so that the value of activity at which we expect to find the crossover between both powerlaws is $|\alpha|=\sigma^2/8K=1/8$. The vertical line in the plot marks this value of $|\alpha|$, while the other two lines are the slopes $q^*\sim |\alpha|$ and $q^*\sim |\alpha|^{1/2}$ that we expect for low and high values of $|\alpha|$ respectively.}
\end{figure}

The expected droplet size as a function of activity in Fig 4 of the main text calculated there as a result of scaling arguments based on the different lengthscales of the system. A more quantitative result can be calculated from matrix $A$ in eqs (5-6) of the main text. In particular, the typical droplet size can be expected to be of a size $L^*\sim 2\pi/q^*$, where $q^*$ is the wavevector at which all higher wavevectors are stable. The stability of different wavevectors is given by the real part of the eigenvalues $E_\pm$ of the matrix $A$. Figure \ref{fig:si_disp} shows the structure of these eigenvalues for a set of parameters where they predict oscillatory states. Solving for the expected droplet size $L^*=2\pi/q^*$ means solving for $E_+=0$, where $E_+$ is the higher eigenvalue. For instance, if we neglect diffusive effects, this eigenvalue is
\eq
E_+=\frac{1}{8} \left(-2 \alpha -4 K q^2+\sqrt{\left(2 \alpha +4 K q^2+q \sigma \right)^2-16 K q^3 \sigma }-q \sigma \right)
\eeq
Figure \ref{fig:qstar} shows the value $q^*$ at which the real part of this eigenvalue changes sign, showing the crossover between two different power laws (in this case $q^*\sim|\alpha|$ for low values of $|\alpha|$ and $q^*\sim|\alpha|^{1/2}$ for high values of $\alpha$) for the expected droplet size in the coexistence region that is shown as the result of numerical simulations in Figure 6 of the main text.

\section{Details on the numerical integration of the equations of motion}
    The numerical integrations of the equations of motion have been done on an OpenFOAM solver, available in the author's github page, along with examples. This solver is based on the PIMPLE OpenFOAM solver for the Navier-Stokes equation, while advancing the equations for $\phi$ and $Q$ with Euler steps. Most of the simulations have been done on lattices of $128\times128$, except for the runs to produce figure 4 of the main text, in which $128$ was not enough to observe enough change in the mean droplet size at the steady state, and sizes of $256\times256$ were used.
    
    We provide videos of two example runs of systems showing arrested phase separation and full mixing.
    
    The first video has parameters $-a=b=1,M = 0.1, \kappa=0.7,\alpha=0$, thus showing a run in the equilibrium case. 
    
    The second has $-a=b=1,M = 0.1, \kappa=0.7, \alpha = -0.1,\lambda=1,K \approx 0,\eta=0.01$, showing an arrested phase separated steady state where hydrodynamics is the main physical mechanism driving the growth of the system.
    
    The third has $a=-0.2,b=1,M=0.1,\kappa=0.7,\alpha=-3.1,\lambda = 1,K=0.09,\eta=0.1$, showing a case when activity and temperature are high enough that the system mixes itself into a uniform state.
    
\bibliography{refs}